\documentclass[10pt,reqno,a4paper]{amsart}
\usepackage{bm,hyperref,color}
\usepackage{graphicx}
\usepackage{amsmath,amssymb}

\advance\textwidth 0.5in \advance\oddsidemargin -0.25in \advance\textheight 0.5in \advance\topmargin -0.5in

    \def\ep{\epsilon}

\def\beea{\begin{eqnarray}}
\def\eea{\end{eqnarray}}

\def\pa{\partial} \def\ti{\tilde}  \def\({\left(} \def\){\right)}

\def\beq{\begin{equation}} \def\ee{\end{equation}} \def\bea{\begin{array}} \def\ea{\end{array}}

\newtheorem{The}{Theorem}
\newtheorem{prop}{Proposition}

\begin{document}
\title {\bf  Integrability Test for Discrete Equations via Generalized Symmetries.}
\author{D. Levi}
\email{levi@roma3.infn.it}
\address{Dipartimento di Ingegneria Elettronica, Universit\`{a} di Roma Tre, and INFN, Sezione di Roma Tre, via della Vasca Navale, 84, Roma, Italy}
\author{ R.I. Yamilov}
\address{Ufa Institute of Mathematics, Russian Academy of Sciences, 112 Chernyshevsky Street, 450008 Ufa, Russian Federation  }


\begin{abstract}
In this article we present some integrability conditions for partial difference equations obtained using the formal symmetries approach.  We apply them to find integrable partial difference equations contained in a class of equations obtained by the multiple scale analysis of the general multilinear dispersive difference equation defined on the square.
\end{abstract}

\maketitle



\section{Introduction}\label{s1}
DL met Marcos Moshinsky for the first time on his arrival in Mexico in February 1973. He went there with a two years fellowship of the CONACYT (Consejo Nacional de Ciencia y Tecnolog\'ia), the Mexican Research Council.
 DL visit was part of his duties as military service in Italy and was following a visit to Mexico  in the summer of 1972 by Francesco Calogero, with whom he graduated at the University of Rome {\it La Sapienza}  in the spring of 1972 with a thesis on "Computation of bound state energy for nuclei and nuclear matter with One-Boson-Exchange Potentials" \cite{calogero}.

The bureaucratic process, preliminary to this visit, was long and tiring and it was successful only for the strong concern of DL uncle, Enzo Levi, professor of Hydrulics at UNAM  \cite{el} and for the help of Marcos Moshinsky who  signed a contract for him  which, as DL uncle wrote,  was ready to substain with his own personal funds if no other source could be found.

During DL stay in Mexico he was able, with the help of Marcos, to appreciate the pleasures of research and to start his career in Mathematical Physics.
With Marcos Moshinsky he published, during the two years stay at the Instituto de Fisica of UNAM, 4 articles \cite{marcos1,marcos2,marcos3,marcos4} but he also collaborated with some of the other physicists of the Institute \cite{brody,dellano1,dellano2}.
This visit was fundamental to shape DL future life, both from the personal point of view and from the point of view of his career.

After 1975 DL met Marcos Moshinsky frequently both abroad and in Mexico. When DL was in Mexico,  Marcos Moshinsky always invited him to his home to partecipate to his family dinners.
As a memento of Marcos, let us add a picture of his with Pavel Winternitz, Petra Seligman and Decio Levi, taken  in September 1999 in front of the Guest House of the CIC-AC where he stayed overnight after  a seminar given at the Instituto de Fisica of UNAM, Cuernavaca branch.

\begin{figure*}
\includegraphics[width=12cm]{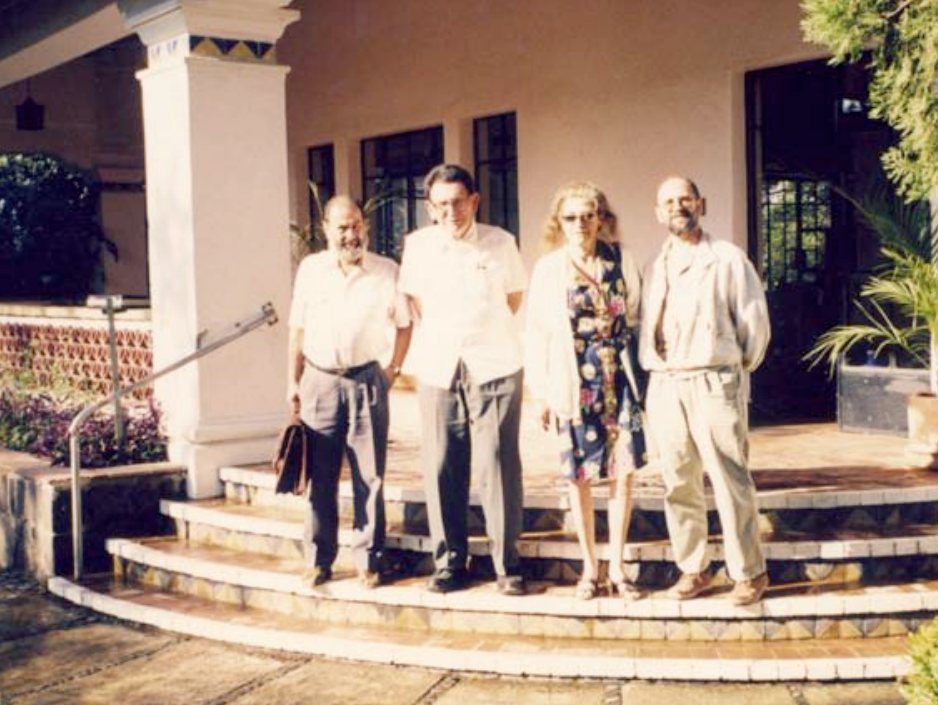}
\caption{\small Pavel Winternitz, Marcos Moshinsky, Petra Seligman and Decio Levi in front of the Guest House of CIC-AC in Cuernavaca in September 1999.}
\label{fig1}
\end{figure*}

Symmetries have played always an important role in physics and in the research of Marcos Moshinsky.  This presentation  shows how crucial can be the notion of symmetry in uncovering integrable structures in nonlinear partial difference equations.

The discovery of  new  { integrable Partial Difference Equations} ($P \Delta E$)
 is always a very challenging problem as, by proper continuous
limits, we can obtain  integrable Differential Difference Equations ($D \Delta E$) and Partial Differential Equations ($PDE$).

A very successful way to uncover integrable $PDE$ has been  the  formal symmetry
approach  due to Shabat and his school in Ufa \cite{shabat}.
These results have been later extended to the case of $D \Delta E$ by  Yamilov \cite{yamilov}, a former student of Shabat.

Here we present some results on the application of the formal symmetry technique to $P \Delta E$.

The basic theory for obtaining symmetries of differential equations has been introduced by Sophus Lie at the end of the nineteen century and can be found together with its extension to generalized symmetries introduced by Emma Noether, for example   in the  book by  Olver  \cite{olver}. The extension of the classical theory to $P \Delta E$ can be found in the work of Levi and Winternitz \cite{lw2006,lysms,wsms}.

Here in the following we outline the results of the geometric classification of $P \Delta E$ given by Adler, Bobenko and Suris \cite{abs} with some critical comments at the end. Then in Section 2 we derive the lowest integrability conditions starting from the request that the $P \Delta E$ admit generalized symmetries of sufficiently high order. In Section 3 we show how the test can be applied. Then  we apply it to a simple class of $P \Delta E$ obtained by  the multiple scale analysis of a generic multilinear dispersive equation defined on the square.

\subsection{ Classification of linear affine discrete equations}
 Adler, Bobenko and Suris \cite{abs} considered the following class of autonomous $P \Delta E$:
\beq\label{i1}
u_{i+1,j+1} = F(u_{i+1,j}, u_{i,j}, u_{i,j+1}) ,
\ee
where $i,j$ are arbitrary integers. Eq. (\ref{i1}) is a discrete analogue of the hyperbolic equations
  \beq\label{i2} u_{xy} = F(u_x, u, u_y) . \ee
which are very important in many fields of physics.  Up to now the general equation (\ref{i2}) has not been classified. Only  the two particular cases: \beea\nonumber u_{xy} = F(u) ;\qquad
 u_x = F(u,v) \;,\;  v_y = G(u,v) , \eea
 which are essentially easier, have been classified by the { formal symmetry approach} \cite{zs1,zs2}.

The ABS  integrable lattice equations are defined as those  autonomous affine linear
(i.e. polynomial of degree one in each argument, i.e. multilinear) partial difference
equations of the form
\beq
\mathcal E (u_{0,0}, u_{1,0}, u_{0,1}, u_{1,1}; \alpha, \beta) = 0, \label{jj}
\ee
where $\alpha$ and $\beta$ are two constant parameters ,
whose {\it integrability} is based on the {\it consistency around a cube} (or 3D-consistency). Here and in the following, as the equations are autonomous, and thus translational invariant, we skip the indices $i$ and $j$ and write the equations around the origin.

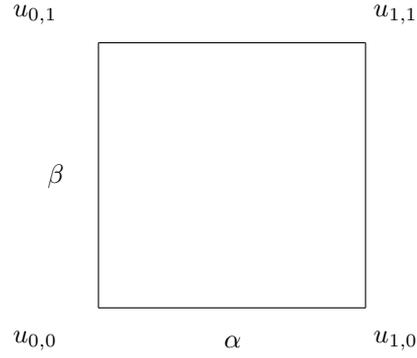
\begin{figure}[htbp]
\setlength{\unitlength}{0.1em}
\begin{picture}(200,140)(-50,-20)
  \put( 0,  0){\line(1,0){100}}
  \put( 0,100){\line(1,0){100}}
  \put(  0, 0){\line(0,1){100}}
  \put(100, 0){\line(0,1){100}}
  \put(-32,-13){$u_{0,0}$}
    \put(-19,47){$\beta$}
     \put(47,-15){$\alpha$}
  \put(103,-13){$u_{1,0}$}
  \put(103,110){$u_{1,1}$}
  \put(-32,110){$u_{0,1}$}
\end{picture}
\caption{A square lattice}
\end{figure}
\begin{figure}[htbp]
\setlength{\unitlength}{0.1em}
\begin{picture}(200,170)(-50,-20)
  \put(100,  0){\circle*{6}} \put(0  ,100){\circle*{6}}
  \put( 50, 30){\circle*{6}} \put(150,130){\circle*{6}}
  \put(  0,  0){\circle{6}}  \put(100,100){\circle{6}}
  \put( 50,130){\circle{6}}  \put(150, 30){\circle{6}}
  \put( 0,  0){\line(1,0){100}}
  \put( 0,100){\line(1,0){100}}
  \put(50,130){\line(1,0){100}}
  \multiput(50,30)(20,0){5}{\line(1,0){15}}
  \put(  0, 0){\line(0,1){100}}
  \put(100, 0){\line(0,1){100}}
  \put(150,30){\line(0,1){100}}
  \multiput(50,30)(0,20){5}{\line(0,1){15}}
  \put(  0,100){\line(5,3){50}}
  \put(100,100){\line(5,3){50}}
  \put(100,  0){\line(5,3){50}}
  \multiput(50,30)(-16.67,-10){3}{\line(-5,-3){12}}
     \put(-10,-13){$u_{0,0,0}$}
     \put(90,-13){$u_{1,0,0}$}
     \put(50,17){$u_{0,0,1}$}
     \put(-13,110){$u_{0,1,0}$}
     \put(160,25){$u_{1,0,1}$}
     \put(45,140){$u_{0,1,1}$}
     \put(109,95){$u_{1,1,0}$}
     \put(157,135){$u_{1,1,1}$}
     \put(40,-13){$\alpha$}
     \put(-16,50){$\beta$}
     \put(20,25){$\gamma$}
\end{picture}
\caption{Three-dimensional consistency}\label{fig.cube}
\end{figure}
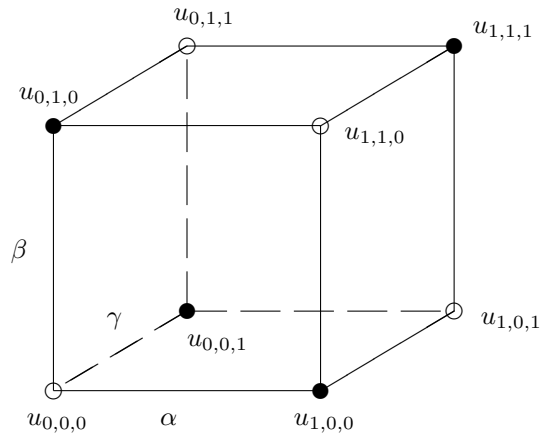

The main idea of the consistency method is the
following:
\begin{enumerate}
\item One starts from a square lattice and defines the three variables $u_{i,j}$ on the
 vertices (see Figure 2).
By solving $\mathcal E = 0$ one obtains a rational expression for the fourth one.
\item One
adjoins a third direction, say $k$, and imagines the map giving $u_{1,1,1}$ as being the composition of maps on the various planes. There
exist {\bf three different ways} to obtain $u_{1,1,1}$ and the  { consistency constraint} is that they all lead to the {\bf same result}.
\item Two further constraints have been
introduced by Adler, Bobenko and Suris:
\begin{itemize}
\item {\it $D_4$-symmetry}:
\beea \nonumber
\mathcal E (u_{0,0}, u_{1,0}, u_{0,1}, u_{1,1}; \alpha, \beta) &=&
 \pm \mathcal E (u_{0,0}, u_{0,1}, u_{1,0}, u_{1,1}; \beta, \alpha)  \\ \nonumber &=&
  \pm \mathcal E (u_{1,0}, u_{0,0}, u_{1,1}, u_{0,1}; \alpha, \beta).
\eea
\item {\it Tetrahedron property}:  $u_{1,1,1}$ is independent of $u_{0,0,0}$.
\end{itemize}
\item The equations are classified according to the following equivalence group:
\begin{itemize}
\item A M\"obius transformation.
\item Simultaneous {\bf point} change of all variables.
\end{itemize}
\end{enumerate}
As a result of this procedure all equations possess a symmetric (in the exchange of the first to the second index)  Lax pair, B\"acklund transformations etc. . Thus the compatible equations are, for all purposes, completely integrable equations.

The ABS list read:
\begin{eqnarray}
{\mbox{{(H1)}}} &\qquad& (u_{0,0}-u_{1,1})\, (u_{1,0}-u_{0,1})\, -\,\alpha \,+ \, \beta \, = \,0 ,\\ \nonumber &&\quad \mbox{The potential discrete KdV equation \cite{lp,nc}}\nonumber \\
{\mbox{{(H2)}}} & & (u_{0,0}-u_{1,1})(u_{1,0}-u_{0,1}) +(\beta-\alpha) (u_{0,0}+u_{1,0}+u_{0,1}+u_{1,1})- \nonumber \\
& & - \alpha^2 + \beta^2 = 0,  \nonumber \\
{\mbox{{(H3)}}} & & \alpha (u_{0,0} u_{1,0}+u_{0,1} u_{1,1}) - \beta (u_{0,0} u_{0,1}+u_{1,0} u_{1,1}) + \delta (\alpha^2-\beta^2) = 0,\nonumber \\
\eea
\beea
{\mbox{{(Q1)}}} & &   \alpha (u_{0,0}-u_{0,1}) (u_{1,0}- u_{1,1}) - \beta (u_{0,0}- u_{1,0}) (u_{0,1} -u_{1,1}) +\nonumber \\
& &    + \delta^2 \alpha \beta (\alpha-\beta)= 0,\quad \mbox{The Schwarzian discrete KdV equation \cite{lps,nqc}}\nonumber\\
{\mbox{{(Q2)}}} & &  \alpha (u_{0,0}-u_{0,1}) (u_{1,0}- u_{1,1}) - \beta (u_{0,0}- u_{1,0}) (u_{0,1} -u_{1,1}) + \nonumber\\
 &&   + \alpha \beta (\alpha-\beta) (u_{0,0}+u_{1,0}+u_{0,1}+u_{1,1}) - \alpha \beta (\alpha-\beta) (\alpha^2-\alpha \beta + \beta^2) = 0,
  \nonumber\\
{\mbox{{(Q3)}}} & &   (\beta^2-\alpha^2) (u_{0,0} u_{1,1}+u_{1,0} u_{0,1}) + \beta (\alpha^2-1) (u_{0,0} u_{1,0}+u_{0,1} u_{1,1})- \nonumber\\
&&    - \alpha (\beta^2-1) (u_{0,0} u_{0,1}+u_{1,0} u_{1,1}) - \frac{\delta^2 (\alpha^2-\beta^2) (\alpha^2-1) (\beta^2-1)}{4 \alpha \beta}=0 , \nonumber \\
{\mbox{{(Q4)}}} & &    a_0 u_{0,0} u_{1,0} u_{0,1} u_{1,1} + \nonumber \\
&&  + a_1 (u_{0,0} u_{1,0} u_{0,1} + u_{1,0} u_{0,1} u_{1,1} + u_{0,1} u_{1,1} u_{0,0} + u_{1,1} u_{0,0} u_{1,0})+ \nonumber \\
& &   +a_2 (u_{0,0} u_{1,1} + u_{1,0} u_{0,1}) + \bar{a}_2 (u_{0,0} u_{1,0}+u_{0,1} u_{1,1})+\nonumber \\
& &  + \tilde{a}_2 (u_{0,0} u_{0,1}+u_{1,0} u_{1,1}) + a_3 (u_{0,0} + u_{1,0} + u_{0,1} + u_{1,1}) + a_4 = 0,\nonumber
\end{eqnarray}
where the seven parameters $a_i$'s in {{(Q4)}} are related by 3 equations.

By a proper limiting procedure all equations of the ABS  list are contained in eq. (Q4) \cite{nah09}.
The symmetries  for the discrete equations of the ABS list have been constructed \cite{ttx,rh} and are given by $D \Delta E$, subcases of Yamilov's discretization of the Krichever--Novikov equation (YdKN) \cite{lpsy,yamilov}:
\beea
 \frac{d u_{0}}{d \ep}= \frac {R(u_{1}, u_0, u_{-1})}{u_1 - u_{-1}} ,
 \quad R(u_{1}, u_0, u_{-1})=A_0 u_1 u_{-1} + B_0 (u_1 + u_{-1}) + C_0 ,
 \nonumber
\eea
where
\beea
&& A_0 = c_1 u_0^2 + 2 c_2 u_0 + c_3, \nonumber  \\
&& B_0 = c_2 u_0^2 + c_4 u_0 + c_5 ,  \nonumber  \\
&& C_0 = c_3 u_0^2 + 2 c_5 u_0 + c_6 .\nonumber
\eea
It is immediate to see that by defining $v_i=u_{i,j}$ and  $\tilde v_i = u_{i,j+1}$, the equations of the ABS list are nothing else but B\"acklund transformations for particular subcases of the YdKN \cite{levi,lpsy}. The ABS equations do not exhaust all the possible B\"acklund transformations for the YdKN equation as the whole parameter space is not covered \cite{lpsy,x09}. Moreover, in the list of integrable $D \Delta E$ of Volterra type \cite{yamilov}, there are equations different from the YdKN which may also have B\"acklund transformations of the form (\ref{i1}). So we have space for new integrable $P \Delta E$ which we will search by using the formal symmetry approach. An extension of the 3D consistency approach has been proposed by the same authors \cite{abs1} allowing different equations in the different faces of the cube. However in this way ABS were able to provide only examples of new integrable $P \Delta E$ but not to present a complete classification scheme.

\section{Construction of Integrability Conditions}\label{s2}

We consider the  class of autonomous $P \Delta E$
 \beq\label{a1} u_{1,1} = f_{0,0} = F(u_{1,0}, u_{0,0}, u_{0,1}) \quad (\pa_{u_{1,0}} F ,\, \pa_{u_{0,0}} F  ,  \, \pa_{u_{0,1}} F ) \ne 0 . \ee
Introducing the
 two shifts operators, $T_1$ and $T_2$ such that  $
T_1 u_{i,j} = u_{i+1,j}, \; T_2 u_{i,j} = u_{i,j+1},
$ it follows that
the functions $u_{i,j}$ are related among themselves by eq. (\ref{a1}) and its shifted values
\beea \nonumber
u_{i+1,j+1} = T_1^i T_2^j f_{0,0} = f_{i,j} = F(u_{i+1,j}, u_{i,j}, u_{i,j+1} ).
\eea
So, the functions $u_{i,j}$ are not all independent. However we can introduce a set of independent functions $u_{i,j}$ in term of which all the others are expressed. A possible choice is given by
$(u_{i,0}, \, u_{0,j})$, for any arbitrary $i,j$  integers.

A generalized symmetry, written in evolutionary form, is given by
\beq\label{a4}
\frac{d}{dt} u_{0,0} = g_{0,0} = G(u_{n,0}, u_{n-1,0}, \dots,
u_{n',0}, u_{0,k}, u_{0,k-1}, \dots, u_{0,k'}) ,\quad n \ge n', \, k \ge k'.
\ee
where $t$ is the group parameter. By shifting, we can write it in any point of the plane
 \[ \frac{d}{dt} u_{i,j} =
T_1^i T_2^j g_{0,0} = g_{i,j} = G(u_{i+n,j}, \dots, u_{i+n',j}, u_{i,j+k}, \dots, u_{i,j+k'}) . \]
In term of the functions $g_{i,j}$ we can write down the symmetry invariant condition
  \beea \label{a4a}
\biggl[g_{1,1} - \frac{df_{0,0}}{dt}\biggr]
 \biggl|_{u_{1,1} = f_{0,0}} = 0 .
\eea
i.e.  $  g_{1,1} = ( g_{1,0} \pa_{u_{1,0}} + g_{0,0} \pa_{u_{0,0}} + g_{0,1} \pa_{u_{0,1}} ) f_{0,0} . $
This equation involves the independent variables $(u_{i,0}, \, u_{0,j})$ appearing in $g_{0,0}$ shifted to points laying on  lines neighboring the axis, i.e. $(u_{i,1}, \, u_{1,j})$. For those function we can state the following Proposition \cite{ly2009}, necessary to prove the subsequent Theorems:

\begin{prop}\label{t3} The functions $u_{i,1},u_{1,j}$ have the following structure:
\beq\label{c9}
\bea{lll} & i>0: \; u_{i,1} = u_{i,1} (u_{i,0},
u_{i-1,0}, \dots, u_{1,0}, u_{0,0}, u_{0,1}) , \quad &\pa_{u_{i,0}} u_{i,1} = T_1^{i-1} f_{u_{1,0}} ;
\\ & i<0: \; u_{i,1} = u_{i,1} (u_{i,0},
u_{i+1,0}, \dots, u_{-1,0}, u_{0,0}, u_{0,1}) , \quad &\pa_{u_{i,0}} u_{i,1} = - T_1^i \frac{f_{u_{0,0}}}{f_{u_{0,1}}} ;
\\ & j>0: \; u_{1,j} =
u_{1,j} (u_{1,0}, u_{0,0}, u_{0,1}, \dots, u_{0,j-1}, u_{0,j}) , \quad &\pa_{u_{0,j}} u_{1,j} = T_2^{j-1} f_{u_{0,1}} ;
\\ & j<0: \; u_{1,j} =
u_{1,j} (u_{1,0}, u_{0,0}, u_{0,-1}, \dots, u_{0,j+1}, u_{0,j}) , \quad &\pa_{u_{0,j}} u_{1,j} = - T_2^j \frac{f_{u_{0,0}}}{f_{u_{1,0}}} . \ea\ee
\end{prop}
In eq. (\ref{c9}) and in the following, $f_{u_{i,j}}=\frac{\partial f_{0,0}}{\partial u_{i,j}}$ and $g_{u_{i,j}}=\frac{\partial g_{0,0}}{\partial u_{i,j}}$.  If a generalized symmetry of characteristic function
$g_{0,0}$ depends on at least one variable of the form $u_{i,0}$,  then $(g_{u_{n,0}},g_{u_{n',0}}) \ne 0$, and the numbers $n,n'$ are called  the {\bf orders
of the symmetry}. The same can be said about the variables $u_{0,j}$ and the corresponding numbers {$k,k'$} if $(g_{u_{0,k}},g_{u_{0,k'}}) \ne 0$.

Now we can state the following Theorem, whose proof can be found in \cite{ly2009}:
\begin{The}\label{t4} If the $P \Delta E$ $u_{1,1}=F$ possesses a generalized symmetry  then
 the following relations must take place: \beq\label{c10}
n>0, \qquad (T_1^n -1) \log f_{u_{1,0}} = (1-T_2) T_1 \log g_{u_{n,0}} ,
\ee\beq\label{c11}
n'<0, \qquad (T_1^{n'} -1) \log \frac{f_{u_{0,0}}}{f_{u_{0,1}}} = (1-T_2) \log g_{u_{n',0}} , \ee\beq\label{c12} k>0, \qquad  (T_2^k -1) \log
f_{u_{0,1}} = (1-T_1) T_2 \log g_{u_{0,k}} , \ee\beq\label{c13}
k'<0, \qquad (T_2^{k'} -1) \log \frac{f_{u_{0,0}}}{f_{u_{1,0}}} = (1-T_1) \log g_{u_{0,k'}} .
\ee \end{The}
 As
 \[\bea{lll}
& T_l^m -1 = (T_l -1) (1 + T_l + \dots + T_l^{m-1}), & m>0, \\ & T_l^m -1 = (1 - T_l) (T_l^{-1} + T_l^{-2} + \dots + T_l^m), & m<0,\quad l=1,2,\ea\]
it follows from Theorem \ref{t4} that we can write the equations (\ref{c10}, \ref{c11}, \ref{c12}, \ref{c13}) as standard conservation laws. Thus, the assumption that a generalized symmetry exist implies the existence of  some conservation laws.

If we assume that a second generalized symmetry exists, i.e. we can find a nontrival function $\tilde G$ such that
\beea \label{c13a}u_{0,0,\ti t} = \ti g_{0,0}=\ti G(u_{\ti n,0}, u_{\ti n-1,0}, \dots,
u_{\ti n',0}, u_{0,\ti k}, u_{0,\ti k-1}, \dots, u_{0,\ti k'}) ,
\eea
where   $\ti n, \ti n', \ti k, \ti k'$ are its orders, then  we can state the following Theorem:
\begin{The}\label{t5} Let the $P \Delta E$ $u_{1,1}=F$ possess two generalized symmetries of orders $(n,n',k,k')$ and $(\tilde n,\tilde n',\tilde k, \tilde k')$, $u_{00,t}=g_{00}$ and $u_{00,\tilde t}=\tilde g_{00}$ , and let their orders
satisfy one of the following conditions: \[\bea{lll} & \hbox{Case 1}: \ n>0, \ \ti n = n+1 & \quad \hbox{Case 2}: \ n'<0, \ \ti n' = n'-1 \\ &
\hbox{Case 3}: \ k>0, \ \ti k = k+1 & \quad \hbox{Case 4}: \ k'<0, \ \ti k' = k'-1 \ea\] Then in correspondence with each of the previous cases  the $P \Delta E$ $u_{1,1}=F$ admits a  conservation law
\beq\label{c14} (T_1 -1) p_{0,0}^{(m)} = (T_2 -1) q_{0,0}^{(m)} , \qquad m=1,2,3,4, \ee
  where  \beq\label{c15} p_{0,0}^{(1)} = \log f_{u_{1,0}} , \qquad p_{0,0}^{(2)} = \log \frac{f_{u_{0,0}}}{f_{u_{0,1}}}
, \qquad q_{0,0}^{(3)} = \log f_{u_{0,1}} , \qquad q_{0,0}^{(4)} = \log \frac{f_{u_{0,0}}}{f_{u_{1,0}}} . \ee \end{The}

So the assumption that the $P \Delta E$ $u_{1,1}=F$ have two generalized symmetries implies that we must have  four necessary conditions of integrability, i.e.  there must exist some functions of finite range $q_{0,0}^{(1)},
q_{0,0}^{(2)}, p_{0,0}^{(3)}, p_{0,0}^{(4)}$ satisfying the conservation laws (\ref{c14}) with $p_{0,0}^{(1)}, p_{0,0}^{(2)},
q_{0,0}^{(3)}, q_{0,0}^{(4)}$ defined by eq. (\ref{c15}). $\;$
 $q_{0,0}^{(1)}$ and $q_{0,0}^{(2)}$ may depend only on the variables $u_{i,0}$, and $p_{0,0}^{(3)}$ and
$p_{0,0}^{(4)}$  on $u_{0,j}$.

Summarizing the results up to now obtained we can say that  a nonlinear partial difference equation will be considered to be integrable if it has a generalized symmetry of finite order, i.e. depending on a finite number of fields.
This provide some conditions which imply the existence of  functions $p_{0,0}^{(m)}$ or $q_{0,0}^{(m)}$ of finite range whose existence is proved by {\it solving a total difference.}

For a $D \Delta E$, when all shifted variables are independent the proof that a total difference has a  solution depending on a finite number of fields, i.e. is a finite range function,  is carried out by applying the discrete analogue of the variational derivative, i.e. a function $q_n$ is (up to a constant) a total difference of a function of finite range iff
\beea
\frac{\delta q_n}{\delta u_n} = \sum_j T^{-j} \frac{\partial q_n}{\partial u_{n+j}} =0,
\eea
see, e.g. \cite{yamilov}.
 For $P \Delta E$ this is no more valid as the shifted variables are not independent as they are related by the  nonlinear $P \Delta E$, in our case  $ u_{1,1} =  F(u_{1,0}, u_{0,0}, u_{0,1})$. This turns out to be the main problem for the application of the formal symmetry approach to $P \Delta E$.

To get a definite result we limit our considerations to
  five points generalized symmetries, i.e. when : \beq\label{d1} \dot
u_{0,0} = g_{0,0} = G (u_{1,0}, u_{-1,0}, u_{0,0}, u_{0,1}, u_{0,-1}) , \quad g_{u_{1,0}} g_{u_{-1,0}} g_{u_{0,1}} g_{u_{0,-1}} \ne 0 . \ee
The existence of a 5 points generalized symmetry will be taken by us as an \textsl{integrability criterion}. This may be a severe restriction as there might be integrable equations with symmetries depending on more lattice points. However just in this case we are able to get sufficiently easily a definite result and, as will be shown in the next Section, we can even solve a classification problem. In this case we can state the following Theorem, which specifies the results obtained so far to the case of five point symmetries:
\begin{The}\label{t7} If the $P \Delta E$ $u_{1,1}=F$ possesses a 5 points generalized symmetry, then the functions
 \beq\label{d2}\bea{lll} & q_{0,0}^{(m)} = Q^{(m)} (u_{2,0}, u_{1,0}, u_{0,0}) , & \quad m=1,2, \\ & p_{0,0}^{(m)} = P^{(m)}
(u_{0,2}, u_{0,1}, u_{0,0}) , & \quad m=3,4, \ea\ee must satisfy the conditions (\ref{c14}, \ref{c15}). \end{The}
Then, using the relations (\ref{c10}--\ref{c13}) with $n=k=1$ and $n'=k'=-1$, we get the following relations between the solutions of the total difference conditions and the generalized symmetry $G$:
\beq\label{d3}\bea{lll} &
q_{0,0}^{(1)} = -T_1 \log G_{,u_{1,0}} , & \qquad
  q_{0,0}^{(2)} =  T_1 \log G_{,u_{-1,0}} , \\
& p_{0,0}^{(3)} = -T_2 \log G_{,u_{0,1}} , & \qquad
  p_{0,0}^{(4)} =  T_2 \log G_{,u_{0,-1}} .
\ea\ee
So, to prove the integrability, which for us means find a generalized 5 point symmetry, for a  nonlinear $P \Delta E$ $u_{11}=F$, we have to check the integrability conditions (\ref{c14}, \ref{c15}). If they are satisfied, i.e. there exist some finite range functions $q_{0,0}^{(m)}$ and $p_{0,0}^{(m)}$,  we can construct   the partial derivatives of $G$. The compatibility of these partial derivatives of $G$,  given by eqs. (\ref{d3}), provides the {\it additional integrability condition}
 \beq\label{d4} G_{,u_{1,0},u_{-1,0}} = G_{,u_{-1,0},u_{1,0}} , \qquad G_{,u_{0,1},u_{0,-1}} = G_{,u_{0,-1},u_{0,1}} . \ee
 If these additional integrability conditions are satisfied, we find $g_{0,0}$ up to an arbitrary
unknown function of the form $\nu(u_{0,0})$, which may correspond  to a Lie point symmetry. This function can be specified, using the determining equations (\ref{a4a}).

The 5 point generalized symmetry $g_{0,0}$, so obtained, will be  of the form:
\beq\label{d5} g_{0,0} = \Phi (u_{1,0}, u_{0,0}, u_{-1,0}) + \Psi
(u_{0,1}, u_{0,0}, u_{0,-1}) + \nu(u_{0,0}). \ee
\section{Application of the test: an example}
To check the integrability conditions (\ref{c14}, \ref{c15}) we need to find the finite range functions $q_{0,0}^{(m)}$ ($m=1,2$) and $p_{0,0}^{(m)}$ ($m=3,4$).  This is not an easy task even if they are linear first order difference equations. A solution  always exists but nothing ensure us a priory that the solution is a finite range function. So let us present a scheme for solving explicitly the integrability conditions we found for the equations on the square i.e. for finding the functions $q_{0,0}^{(1)}$,  $q_{0,0}^{(2)}$, $p_{0,0}^{(3)}$ and $p_{0,0}^{(4)}$.

As an example of this procedure let us consider the solution of eq. (\ref{c14}) for $m=1$, where
\beea \label{f1}
p_{0,0}^{(1)} =\log(f_{u_{1,0}}), \; q_{0,0}^{(1)} = Q^{(1)} (u_{2,0}, u_{1,0}, u_{0,0}), \;T_2 q_{0,0}^{(1)} = Q^{(1)} (u_{2,1}, u_{1,1}, u_{0,1}).
\eea
In eq. (\ref{f1}) we have the dependent variables $u_{2,1}$ and $u_{1,1}$ where
$u_{2,1}=F(u_{2,0},u_{1,0},u_{1,1})$ while $u_{1,1}=F(u_{1,0},u_{0,0},u_{0,1})$. So  eq.  (\ref{c14}) for $m=1$ will contain the unknow function $F$ which characterize the class of equations we are considering twice, one time to calculate $u_{1,1}$ in terms of independent variables and then to calculate $u_{2,1}$ in term of $u_{1,1}$ and of the independent variables. This double dependence makes the calculations extremely difficult. To overcome this difficulty we take into account that we are considering autonomous equations which are shift invariant. So we can substitute  eq.  (\ref{c14}) for $m=1$ with the following equivalent independent equations
\beea \label{f2}
&p_{0,0}^{(m)}-p_{-1,0}^{(m)} &= Q^{(m)} (u_{1,1}, u_{0,1}, u_{-1,1})-Q^{(m)} (u_{1,0}, u_{0,0}, u_{-1,0})\\ \label{f3}
&p_{0,-1}^{(m)}-p_{-1,-1}^{(m)} &= Q^{(m)} (u_{1,0}, u_{0,0}, u_{-1,0})-Q^{(m)} (u_{1,-1}, u_{0,-1}, u_{-1,-1})
\eea
where, to simplify the notation, we introduce in the following the functions
\beea \nonumber
&u_{1,1}=f^{(1,1)}(u_{1,0}, u_{0,0},u_{0,1}),\quad &u_{-1,1}=f^{(-1,1)}(u_{-1,0}, u_{0,0},u_{0,1}), \\ \nonumber
&u_{1,-1}=f^{(1,-1)}(u_{1,0}, u_{0,0},u_{0,-1}),\quad &u_{-1,-1}=f^{(-1,-1)}(u_{-1,0}, u_{0,0},u_{0,-1}),
\eea
to indicate $f_{0,0}$ and its analogues.
Moreover, we  introduce the following two differential operators
\beea \label{f5}
\mathcal A &=& \partial_{u_{0,0}} - \frac{f^{(1,1)}_{u_{0,0}}}{f^{(1,1)}_{u_{1,0}}}  \partial_{u_{1,0}}- \frac{f^{(-1,1)}_{u_{0,0}}}{f^{(-1,1)}_{u_{-1,0}}}  \partial_{u_{-1,0}},\\ \nonumber
\mathcal B &=& \partial_{u_{0,0}} - \frac{f^{(1,-1)}_{u_{0,0}}}{f^{(1,-1)}_{u_{1,0}}}  \partial_{u_{1,0}}- \frac{f^{(-1,-1)}_{u_{0,0}}}{f^{(-1,-1)}_{u_{-1,0}}}  \partial_{u_{-1,0}}.
\eea
 in such a way  that the functional equations (\ref{f2}, \ref{f3}) reduce to  {\it differential monomials} \cite{abel}:
\beea \label{f4}
&\mathcal A Q^{(m)} (u_{1,1}, u_{0,1}, u_{-1,1}) =0, \qquad &\mathcal B Q^{(m)} (u_{1,-1}, u_{0,-1}, u_{-1,-1})=0, \\
\label{f6}
&\mathcal A Q^{(m)} (u_{1,0}, u_{0,0}, u_{-1,0}) =r^{(m,1)}, \qquad &\mathcal B Q^{(m)} (u_{1,0}, u_{0,0}, u_{-1,0}) =r^{(m,2)}.
\eea
Eqs. (\ref{f4}) are, by construction, identically satisfied while eqs. (\ref{f6}) provide a set of equations for the derivatives of $Q^{(m)} (u_{1,0}, u_{0,0}, u_{-1,0})$ with respect to its three arguments.  By commuting the two operators (\ref{f5}) we can obtain a third equation for the derivatives of $Q^{(m)} (u_{1,0}, u_{0,0}, u_{-1,0})$ with respect to its three arguments:
\beea \label{f7}
 \qquad \qquad [\mathcal A, \mathcal B]\, Q^{(m)} (u_{1,0}, u_{0,0}, u_{-1,0}) =r^{(m,3)}.
\eea
Eqs. (\ref{f6}, \ref{f7}), if independent, define uniquely  the derivatives of the function $Q^{(m)} (u_{1,0}, u_{0,0}, u_{-1,0})$ and, if their consistency is satisfied, from them we get the functions themselves.

In a similar manner from $(T_1-1)p_{0,0}^{(m)} = (T_2-1)q_{0,0}^{(m)}$ with $m=3,4$ we get the function
$p_{0,0}^{(m)} = P^{(m)} (u_{0,2}, u_{0,1}, u_{0,0})$ and consequently the symmetry (\ref{d5}).

This procedure works if the function $F$ is known, i.e. if we check a given equation for its integrability. It also works if $F$ is known up to some unknown arbitrary constants to be specified. In such case we solve a classification problem with unknown constants. However, the problem is much more difficult if $F$ depends on unknown arbitrary functions of one, two or three variables. In such a case  the coefficients of the operators (\ref{f5}) and functions $r^{(m,k)}$ will depend on unknown functions, and $r^{(m,k)}$ may even depend on the composition of unknown functions. In this case a more complicated procedure might be necessary.
\subsection{A concrete example}
Let us consider the following $P \Delta E$ \cite{hls}
\beea \label{g1}
2(u_{0,0} + u_{1,1}) &+& u_{1,0} + u_{0,1}+
\gamma[4 u_{0,0} u_{1,1} + 2 u_{1,0} u_{0,1} + 3 (u_{0,0} + u_{1,1} )(u_{1,0} + u_{0,1} )]+ \\ \nonumber
&+&(\xi_2 + \xi_4 )u_{0,0} u_{1,1} (u_{1,0} + u_{0,1} ) + (\xi_2 - \xi_4 )u_{1,0} u_{0,1} (u_{0,0} + u_{1,1} )+ \\ \nonumber &+&
\zeta u_{0,0} u_{1,1} u_{1,0} u_{0,1} = 0 .
\eea
Eq. (\ref{g1}) is a dispersive multi--linear partial difference equation which passes the $A_3$ multiple scales integrability test \cite{hl}.
Applying the M\"obious transformation $u_{i,j} = 1/(\hat u_{i,j} - \gamma)$ we can rewrite it in a simplified form as
\beea \label{g2}
(u_{0,0} u_{1,1} + \alpha)(u_{1,0} + u_{0,1} ) + (2u_{1,0} u_{0,1} + \beta)(u_{0,0} + u_{1,1} ) + \delta = 0,
\eea
where $\alpha$, $\beta$ and $\delta$ are well defined functions of $\gamma$, $\xi_2$, $\xi_4$ and $\zeta$.
We now apply to eq. (\ref{g2}) the procedure outlined at the beginning of this section. Eq. (\ref{g2}) depends on three free parameters and we look for conditions on the three parameters, if any, such that the equation admits generalized symmetries. We get that the conditions are satisfied only in two cases:
\begin{enumerate}
 \item $\alpha=2\beta\neq 0$, $\delta=0$ and, as $\beta \neq 0$ we can always set $ \beta=1$. This choice of the parameters  $\alpha$ and $\beta$ corresponds to  $\xi_2 =3 \xi_4 + 3 \gamma^2$ and $\zeta = 12 \gamma \xi_4$ in eq. (\ref{g1}). The corresponding  integrable $P \Delta E$ reads:
\beea \label{g3}
(u_{0,0} u_{1,1} + 2)(u_{1,0} + u_{0,1} ) + (2u_{1,0} u_{0,1} + 1)(u_{0,0} + u_{1,1} ) = 0 .
\eea
In correspondence with the eq. (\ref{g3}) we get the generalized symmetry
\beea \label{g4}
 u_{0,0;t} &=& \frac{(u_{0,0}^2 - 2)(2 u_{0,0}^2-1)}{u_{0,0}} \left\{ A \left[\frac{1}{u_{1,0} u_{0,0} +1} - \frac{1}{ u_{-1,0}u_{0,0} +1}\right]\right. + \\ \nonumber &+& \left.B \left[\frac{1}{u_{0,1} u_{0,0} +1} - \frac{1}{u_{0,-1} u_{0,0}+1}\right] \right\} .
\eea
\item $\beta= 2 \alpha\neq 0$, $\delta=0$ and, as $\alpha \neq 0$ we can always set  $\alpha=1$. This choice of the parameters  $\alpha$ and $\beta$  corresponds to  $\xi_2 =  6 \gamma^2 - 3 \xi_4$ and $\zeta = 12 \gamma (\gamma^2 - \xi_4)$ in eq. (\ref{g1}).  The corresponding  integrable $P \Delta E$ reads:
\beea \label{g5}
(1 + u_{0,0} u_{1,1} )( u_{1,0} + u_{0,1} ) +2 (1 + u_{0,1} u_{1,0} )( u_{0,0} + u_{1,1} )=0.
\eea
In correspondence with the eq. (\ref{g5}) we get the generalized symmetry
\beea \label{g6}
u_{0,0;t}=A(u_{0,0}^2 -1)\frac{u_{1,0}-u_{-1,0}}{u_{-1,0} u_{1,0}-1}+ B(u_{0,0}^2 -1)\frac{u_{0,1}-u_{0,-1}}{u_{0,-1} u_{0,1}-1}
\eea
\end{enumerate}
Here $A,B$ are constant coefficients, and in both cases, ($A=0,B\ne 0$) and ($A\ne 0,B=0$), the nonlinear $D \Delta E$  (\ref{g4}, \ref{g6}) are, up to a point transformation,   equations belonging to the classification of Volterra type equations done by Yamilov \cite{yamilov}. This shows that the eqs. (\ref{g3}, \ref{g5}) do not belong to the ABS classification.
Moreover this calculation shows that the $A_3$ integrability in the multiple scale integrability test is not sufficient to select integrable $P \Delta E$ on the square having five points generalized symmetries.
\section*{Acknowledgments.} R.I.Y. has been partially supported by the Russian Fo\-undation for Basic Research (Grant number 10-01-00088-a and 08-01-00440-a). LD  has been partly supported by the Italian Ministry of Education and Research, PRIN
``Nonlinear waves: integrable finite dimensional reductions and discretizations" from 2007
to 2009 and PRIN ``Continuous and discrete nonlinear integrable evolutions: from water
waves to symplectic maps" from 2010.

\end{document}